\documentclass[12pt]{iopart}

\usepackage{graphicx}

\def\TReg{\textsuperscript{\textregistered}}

\begin{document}

\title[The Tayler instability at low magnetic Prandtl numbers]{The Tayler instability at low magnetic Prandtl numbers: 
between chiral symmetry breaking and helicity oscillations}

\author{Norbert Weber, Vladimir Galindo, Frank Stefani, and Tom Weier }
\address{Helmholtz-Zentrum Dresden - Rossendorf, 
P.O. Box 510119, 01314 Dresden, Germany}
\ead{Norbert.Weber@hzdr.de}
\begin{abstract}
The Tayler instability is a kink-type, current driven 
instability that plays an important role in plasma 
physics
but might also be relevant in liquid metal 
applications with
high electrical currents. In the framework of the
Tayler-Spruit dynamo model of stellar magnetic field
generation \cite{Spruit2002}, the question of spontaneous 
helical (chiral) symmetry breaking during the saturation of the 
Tayler instability
has received considerable interest 
\cite{Zahn2007,Gellert2011,Bonanno2012}.
Focusing on fluids with low magnetic Prandtl numbers, 
for which the quasistatic approximation can be applied,
we utilize an  integro-differential equation 
approach \cite{Weber2013} in order to investigate 
the saturation mechanism of the Tayler instability. 
Both the exponential growth phase and the saturated 
phase are analyzed in terms of the action of 
the $\alpha$ and $\beta$ effects of mean-field
magnetohydrodynamics. In the exponential growth phase
we always find a spontaneous chiral symmetry breaking which, 
however, disappears in the saturated phase. For higher 
degrees of supercriticality, we observe helicity 
oscillations in the saturated regime.
For Lundquist numbers in the 
order of one we also obtain chiral symmetry 
breaking of the saturated magnetic field.

\end{abstract}

\submitto{\NJP}
\maketitle

\section{Introduction} % Motivation

Electrical currents through an incompressible, viscous and 
resistive liquid conductor produce
azimuthal magnetic fields which, beyond a critical field strength,
become unstable to a non-axisymmetric, i.e.~kink-type instability 
that we will call Tayler instability (TI) 
as a tribute to the seminal contributions of R.J. Tayler
\cite{Tayler1957,Tayler1973}. 

For a constant current density in an infinitely long 
cylinder R\"udiger \etal  had shown  \cite{Ruediger2011,Ruediger2013} 
that the governing parameter 
is the Hartmann number, 
$Ha=B_{\varphi}(R) R (\sigma/\rho \nu)^{1/2}$, 
which has to exceed a value in the order of 20 for the
TI to set in ($B_{\varphi}(R)$ is the azimuthal field at 
the outer radius $R$ of the
cylinder,  $\sigma$, $\rho$ and $\nu$ are the
conductivity, density and viscosity of the fluid, respectively).
This critical value of $\mathit{Ha}$
is actually consistent with previous results  
\cite{Spies1988,Shan1991,Montgomery1993,Cochran1993}
concerning the effects of viscosity and resistivity on 
the stability of various plasma z-pinches, 
if one leaves aside the effects of complicated 
boundary conditions and
non-homogeneous material parameters in the plasma case. 
Note that at these early days \cite{Montgomery1993}
it was  far from obvious that the 
governing parameter for the onset of the instability is
$Ha$, rather than the 
Lundquist number $S=Ha Pm^{1/2}$
(with $Pm=\nu \mu_0 \sigma$ denoting the 
magnetic Prandtl number, where 
$\mu_0$ is the magnetic permeability constant).

Whilst the focus of fusion related pinch experiments was 
prominently on the plasma destabilization  when 
the ratio of axial to azimuthal magnetic field 
(the so-called safety parameter) 
falls below a certain critical value
\cite{Bergerson2006}, a recent liquid 
metal experiment, with 
uniform conductivity and viscosity as well as 
well-defined insulating 
boundary condition, 
has indeed confirmed the TI-threshold of $\mathit{Ha}\simeq 20$ 
\cite{Seilmayer2012}. From  the application point of view,
current-driven instabilities in liquid metals 
are presently  considered a possible limitation 
for the integrity 
of large-scale liquid metal batteries. 
Such batteries are self-assembling stratified systems 
made of a heavy liquid metal or metalloid (e.g., Bi, Sb) 
at the bottom, a suitable
molten salt mixture as electrolyte in the middle, and a light
alkaline or earth alkaline metal (e.g., Na, Mg) at the
top. While small versions of liquid metal batteries have 
already been tested \cite{Weaver1962,Cairns1969,Bradwell2012,Kim2012}, the 
occurrence of the TI could possibly present a serious problem for 
the stratification in larger batteries with prospected
charging/discharging currents of some thousand amps. 
In \cite{Stefani2011} we had advised a
simple trick to suppress the TI in liquid metal batteries by
just returning the battery current through a bore in the
centre. By the resulting change of the radial profile
$B_{\varphi}(r)$ it is possible to avoid the (ideal)
condition $\partial (r B^2_{\varphi}(r))/\partial r>0$  
\cite{Tayler1973} for the onset of the TI.

In a follow-up paper \cite{Weber2013}, a numerical code has 
been presented that is capable of treating TI-problems at small
values of 
$Pm$ as they are typical for liquid metals.
This was achieved by replacing the solution of the
induction equation for the 
magnetic field by applying the so-called quasistatic 
approximation \cite{Davidson2001}. This approximation allows to 
avoid the explicit time stepping of the magnetic field by 
computing the electrostatic potential by a Poisson equation,
and deriving from this the electric current density.
The induced magnetic field is then computed from the 
induced current density 
via Biot-Savart's law. This way one arrives at an 
integro-differential
equation approach, as it had already been used 
by Meir and Schmidt for different magnetohydrodynamic (MHD)
problems \cite{Meir2004}. 
Our numerical scheme utilizes the open source CFD library
OpenFOAM\TReg\ \cite{openfoam},  supplemented by an 
MPI-parallelized
implementation of Biot-Savart's law. 
This code was then applied to a number 
of TI related 
problems, in particular for determining the scaling properties
of the growth rate and the saturated velocity field, the dependence
of the critical current on the geometric aspect ratio, as well
as for validating various methods of preventing
TI in liquid metal batteries  \cite{Weber2014}.
Recently, our results were confirmed by 
another code working completely in the framework of
the differential equation approach, by 
analyzing the scaling properties of the solutions
with $Pm$  \cite{Herreman2015}. The authors also 
discussed carefully the limitations of the quasistatic 
approach
for higher values of $Pm$.

An interesting by-product of the battery-oriented 
simulations \cite{Weber2013} was the 
observation of the transient occurrence, but ultimate  
disappearance, of helical structures during the evolution 
of the TI. On the first glance, the appearance of helical 
structures is surprising, since the underlying equations 
have no preference for left or right handed solutions.
Yet, it is exactly this helical (or chiral) symmetry 
breaking that has gained considerable interest in various
astrophysical problems. This applies in particular to the 
concept of the Tayler-Spruit dynamo \cite{Spruit2002} 
in which an azimuthal magnetic field is thought to 
become strong enough to drive the TI against the stable 
stratification in the radiation zone 
of a star. Combined with the usual differential
rotation this effect might lead to a working 
dynamo. Despite of the  attractiveness of the TI, in particular
for explaining angular momentum transport in various types of stars
\cite{Ruediger2013,Meynet2011,Maeder2014},
the concept of the actual Tayler-Spruit dynamo is not without
caveats. Zahn \etal \cite{Zahn2007} 
have argued that the TI-produced
non-axisymmetric ($m=1$) poloidal magnetic field alone would not 
be suited to close the dynamo loop (since the toroidal
field wound up from it would have the same $m=1$ dependence),
but that some sufficiently large mean-field 
$\alpha$ effect would be needed
to produce the necessary axisymmetric poloidal field.

It is exactly here where the question whether TI saturates 
with a finite helicity, produced by a finite $\alpha$ effect, 
becomes highly relevant. Some recent papers have answered this question 
affirmatively: Gellert \etal \cite{Gellert2011} have
found spontaneous chiral symmetry breaking
of the TI in simulations with $Pm$ of  0.1, 1, and 10.
Bonanno \etal \cite{Bonanno2012} got a similar result 
for very  
large $Pm=10^7$. In addition to the numerical simulation, 
the latter authors developed a simple 
model of energy and helicity evolution resulting in       
an instructive phase portrait.                   
The equations describing this behaviour
can also be linked to a similar 
chiral symmetry breaking in biochemistry 
where it refers to the selection of one of two
possible forms of bio-molecules (mainly sugars and 
amino acids) that are mirror images of each other 
\cite{Saito2013}.

With this background, the main motivation for the 
present paper is the discrepancy between 
the simulations of 
\cite{Gellert2011,Bonanno2012,Chatterjee2011}
and the preliminary result
of our low $Pm$ simulations \cite{Weber2013} 
showing that helicity starts to grow but ultimately decays 
to zero. Given the different $Pm$ at which the 
respective simulations were done, it is worthwhile to 
understand in detail the saturation mechanism of TI in 
dependence on $Pm$.

Actually, helical states have a long history in plasma 
physics, tracing back to the early work of Lundquist  
on ''Magneto-hydrostatic fields'' \cite{Lundquist1950}. 
Specializing general pressure-balanced fields to  
force-free fields that satisfy $(\nabla \times {\bi{B}}) \times {\bi{B}}=0$,
he found, first, that fields with $\nabla \times {\bi{B}}= a(r) {\bi{B}}$ fulfill
this demand, and second, that $a(r)=const$ must be 
requested for the field to remain force-free during 
its time-evolution. 
For cylindrical geometry, Lundquist found 
that the force-free condition, i.e.~the 
demand that the current 
is parallel to the field, is guaranteed by Bessel 
function profiles 
$B_z =A J_0(a r), B_{\varphi}=A J_1(a r)$ (interestingly, 
the very same profiles for the {\it velocity} field 
turned later out to provide the most
efficient dynamo of the Ponomarenko or Riga type 
\cite{Stefani1999}).

Soon after Lundquist's work, Chandrasekhar and 
Woltjer \cite{Chandrasekhar1958} interpreted 
this Bessel functions solution in terms of achieving 
''maximum magnetic energy for a given mean-squared 
current density'' or, alternatively, as a 
''state of minimum dissipation for a given magnetic energy''. 
Since Bessel functions also maximize 
the magnetic helicity for given magnetic energy (and  
magnetic helicity is a better conserved quantity 
than energy) a surge of work was devoted to 
understand how solutions of this kind can be 
achieved dynamically. 
This goes mainly under the notion of Taylor relaxation 
\cite{Taylor1986}, and has found great interest in 
connection with the reversed field pinch.
Quite a number of workers have tried to understand Taylor
relaxation from different thermodynamic principles, 
such as minimum entropy production  
\cite{Hameiri1987} or minimum dissipation rates 
\cite{Montgomery_Phillips1988,Montgomery_Phillips_Theobald1989,
Farengo1995,Phillips1996,
Bhattacharyya2001,Dewar2008,Dasgupta2009}.

One of the first applications  of 
a general thermodynamic principle to plasma relaxation 
goes back to a
note  of Max Steenbeck \cite{Steenbeck1932} (relying, in turn, 
on an earlier idea of Compton and Morse  \cite{Compton1927}). 
Steenbeck's principle 
states that in  real gas discharges at fixed current the heat power, 
and thus the voltage drop between the electrodes, 
is minimized (somewhat surprisingly, 
this minimum-dissipation principle
corresponds perfectly with  the maximum entropy production 
rate principle \cite{Martyushev2006}
if one considers the {\it total system} including the 
current-stabilizing external resistor \cite{Christen2009}.)

Interestingly, it was also Steenbeck
who was later to create the theoretical
framework that nowadays allows for a deeper {\it dynamical}
understanding of those somewhat vague thermodynamic 
principles. Mean-field magnetohydrodynamics (MHD) was 
originally developed to explain
self-excitation of cosmic magnetic fields
\cite{Steenbeck1966}. 
Its main idea is that certain correlations
of the  small-scale parts of velocity and magnetic 
field contribute to the dynamics of the large 
scale magnetic field \cite{Krause1980}.
For helical turbulence 
the authors introduced the celebrated $\alpha$ effect 
which drives an electromotive force parallel to 
a prevailing large scale magnetic field $\overline{\bi{B}}$. 
Similarly, turbulence  leads to an
increase of the resistivity by the $\beta$ 
effect, so that the mean electromotive force can be written
in the form  $\cal{E}=\alpha \overline{\bi{B}} -\beta \nabla \times \overline{\bi{B}}$.

Nowadays, mean-field concepts play
not only a role in dynamo theory but also in the description of
magnetically driven instabilities. Flow-driven helical dynamos 
and magnetically dominated helical ``dynamos'' are 
presently considered as two different aspects of
the very same mean-field MHD 
\cite{Blackman2006}.
The detailed  saturation mechanism of the TI, 
in particular its termination in a 
helical or non-helical state, is but one
interesting application of mean-field MHD.

In this paper, we are going to study the exponential
growth and the final saturation of the TI in finite
cylindrical geometry for varying values of $Ha$ and $Pm$.
On the basis of an axisymmetric ($m=0$) base state with
an homogeneous axial current $J_0$ that produces an azimuthal
magnetic field $B_0$, we compute the $m=1$ TI-eigenmode
comprising the velocity $\bi u$ and the induced 
magnetic field $\bi b$ from which we  
infer the mean electromotive force
$\overline{  {\bi{u}} \times {\bi{b}}}$ (the overbar denotes the
average over the azimuthal angle), and from this the mean-field
coefficients $\alpha$ and $\beta$. As a product of 
two $m=1$ modes,   ${\bi{u}} \times {\bi{b}}$   comprises 
certain $m=0$ components that drive an azimuthal 
current (by virtue of the $\alpha$ effect) and
reduce the impressed 
axial current (by virtue of the $\beta$ effect).
Although there is not a big scale separation between 
the $m=0$ base state and the $m=1$  
perturbation, mean-field theory perfectly applies here.
The electromotive force $\cal{E}$ in direction of the
mean field $\overline{\bi{B}}$ (i.e. the $\alpha$ effect), 
will be interpreted in terms of its relation to the 
small-scale current helicity,
$\cal{E} \cdot \overline{\bi{B}}=-\overline{ \bi{j} \cdot \bi{b}}/\sigma+
\overline{\bi{e} \cdot \bi{b}}$,
that had been derived and utilized by different authors
\cite{Bhattacharjee1986,Seehafer1994,Ji1999,Blackman2007,Ebrahimi2014}.

In case of $S \gg 1$, the modified currents and fields
could be expected to resemble the typical Bessel function
structure as typical for Taylor relaxation.
In this sense, the mean axial field produced by the 
$\alpha$ effect would follow from the principle of 
minimum dissipation \cite{Montgomery_Phillips_Theobald1989}.

However, this type of saturation mechanism, which relies 
on changing - by mean-field induction effects - the electromagnetic 
base state in such a way that it becomes just marginally 
stable against TI, does not apply for $S\ll 1$. In this case 
the magnetic Reynolds number  $Rm$ of the TI-produced flow 
is much too small
to induce any significant changes of the
original applied magnetic field. The saturation
must instead rely on a modification of the {\it hydrodynamic}
base state, which we will discuss in detail. We
will also evidence the occurrence of helicity oscillations, 
whose amplitudes and frequencies in dependence on
$Ha$ and $Pm$ we will characterize.

The paper closes with a discussion of the results, and 
with an outlook towards an application to stellar dynamo theory.

\section{The numerical scheme}

The usual numerical schemes for the simulation of TI, 
which solve the Navier-Stokes equation for the velocity and 
the induction equation for the magnetic field, are working 
typically only for values of  $Pm$ down to 10$^{-3}$,
although in a recent work  by Herreman \etal this limit has 
been challenged \cite{Herreman2015}.

Here, we circumvent the usual $Pm$ limitations of 
these codes by replacing the solution of the
induction equation for the 
magnetic field by invoking the so-called quasistatic 
approximation \cite{Davidson2001}. We replace the 
explicit time stepping of 
the magnetic field by computing the electrostatic 
potential by a Poisson equation,
and deriving the electric current density.
However, in contrast to many other inductionless approximations
in which this procedure is sufficient,
in our case we cannot avoid to compute the induced 
magnetic field, too. The reason for that is the 
presence of an externally applied electrical current
in the fluid. Computing the Lorentz force term it turns out that
the product of the applied current with the induced field 
is of the same
order as the product of the magnetic field 
(due to the applied current) with the induced current.
Here, we compute the induced magnetic field 
from the induced current density
by means of Biot-Savart's law. This way we arrive at an 
integro-differential
equation approach, as it had already been used by 
Meir and Schmidt \cite{Meir2004}. 

In detail, the numerical model as developed by Weber \etal 
\cite{Weber2013} works as follows: it solves the Navier-Stokes 
equations (NSE) for incompressible fluids
\begin{eqnarray}\label{eqn:navierstokes}
\dot {\bi u} + \left({\bi u}\cdot\nabla\right){\bi u} = 
- \nabla p + \nu \Delta {\bi u} + \frac{\bi f_{\mathrm
    L} }{\rho}\hspace{5mm}\textrm{and}\hspace{5mm} \nabla\cdot \bi u = 0,
\end{eqnarray}
with $\bi u$ denoting the velocity, $p$ the (modified) pressure, 
$\bi f_{\mathrm L} = \bi J \times \bi B $ the
electromagnetic Lorentz force density, $\bi J$ the 
total current density and $\bi B$
the total magnetic field. The NSE
is solved using the PISO algorithm and applying no slip boundary
conditions at the walls. 

Ohm's law in moving conductors
\begin{eqnarray}
{\bi j} = \sigma\left(-\nabla\varphi + {\bi u}\times {\bi B}\right)
\end{eqnarray}
allows to compute the induced current $\bi j$ by previously
solving a Poisson equation for the perturbed electric potential
$\varphi = \phi -J_0z/\sigma$:
\begin{eqnarray} \Delta\varphi = 
\nabla\cdot\left({\bi u} \times {\bi B}\right).
\end{eqnarray}
In the following, we will concentrate on cylindrical 
geometries with an axially
applied current. Then, after 
subtracting the (constant) potential part 
$J_0z/\sigma$, with $z$ as
coordinate along the cylinder axis, we use the simple
boundary condition $\varphi = 0$ on top and bottom and $\bi n\cdot \nabla
\varphi=0$ on the mantle of the cylinder, with $\bi n$ as the surface
normal vector. 

The induced magnetic field can then be
calculated by Biot-Savart's law
\begin{eqnarray}\label{eqn:biotsavart}
{\bi b}({\bi r}) = \frac{\mu_0}{4\pi}\int dV' \, 
\frac{{\bi j}({\bi r}') \times ({\bi r}-{\bi r}')}{\left|{\bi r}-{\bi r}'\right|^3}.
\end{eqnarray}
Since this is a costly procedure, we modify here the 
method of \cite{Weber2013} slightly. Actually, equation (4)
is applied only at the boundary of the cylinder, while
the magnetic field in the bulk is computed by solving
the vectorial Poisson equation 
$\Delta {\bi b}=\mu_0 \sigma \nabla \times ( {\bi{u}} \times {\bi{B}} )$
which results from the full time-dependent induction equation
in the quasi-stationary approximation.

\begin{figure}[h]
\centerline{
\includegraphics[width=0.7\columnwidth]{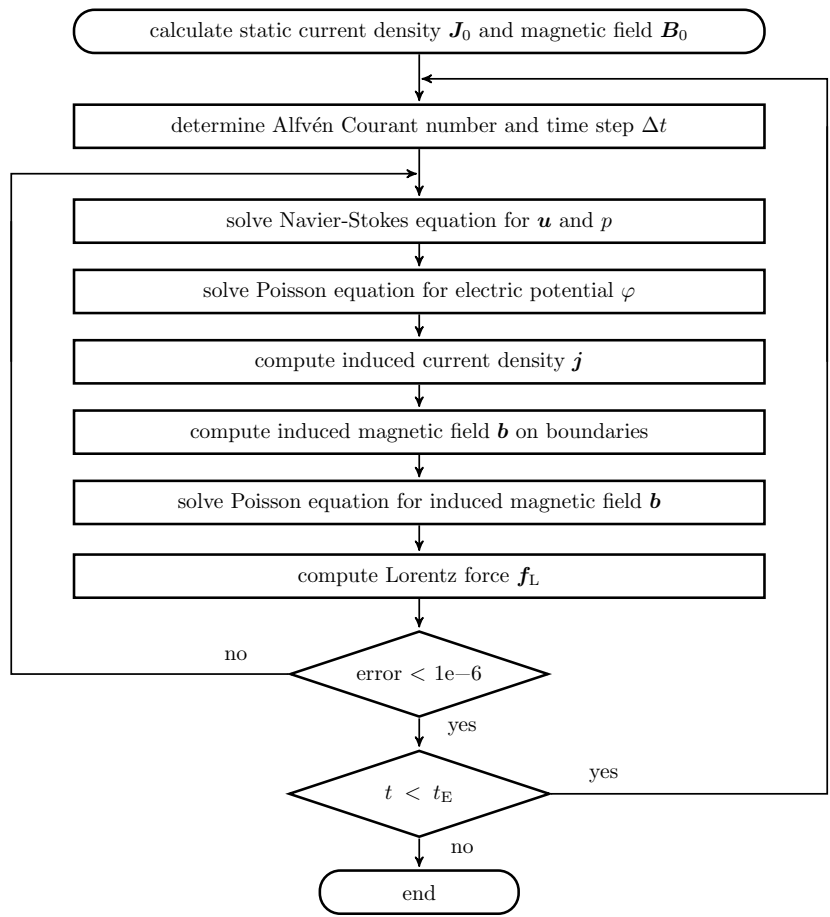}}
\caption{Flow chart of the simulation model, slightly modified 
with respect to that of \cite{Weber2013}.}
\label{fig:fig1}
\end{figure}

Knowing $\bi b$ and $\bi j$ we compute the
Lorentz force ${\bi f}_{\mathrm L}$ for the next iteration. 
A flow chart of this numerical
procedure is shown in figure \ref{fig:fig1}. For more 
details about the numerical scheme, see section 
2 and 3 of \cite{Weber2013}.

\section{Results}

For the sake of simplicity, we 
consider a cylindrical electrically conducting
fluid with a ratio of height $L$ to diameter $2R$ 
of 1.25. A current is applied from top to bottom, just 
by setting the electrical potential
to constant (but different) values at the two 
faces. Note that we refrain from
taking into account any currents in the
electrodes at top and bottom which has been
shown to lead only to minor modifications of
the results \cite{Weber2015}.
The side walls of the cylinder are considered as electrically
insulating. No-slip boundary conditions apply to the velocity 
at all boundaries.

In the following, we will focus on three different 
cases. The magnetic 
Prandtl number for the first two runs is $Pm=10^{-6}$.
Differing in the Hartmann number (60  and 100), these two 
runs will show a quite different behaviour of the helicity. 
Whereas for $Ha=60$ the initially growing helicity 
ultimately  goes to zero, for $Ha=100$ 
we observe helicity oscillation 
in the saturation regime.
For the third case, with the much higher  
$Pm=10^{-3}$ and $Ha=100$, 
we  will find a finite and non-zero value of the 
final helicity. 
At the end of this section, we will summarize the 
different ways of saturation.

\subsection{Saturation with zero helicity}

Here, we choose $Ha=60$ and $Pm=10^{-6}$
which results in a Lundquist number $S=0.06$ that 
is definitely low enough for 
applying the  quasistatic approximation.

\begin{figure}[h]
\centerline{
\includegraphics[width=0.99\columnwidth]{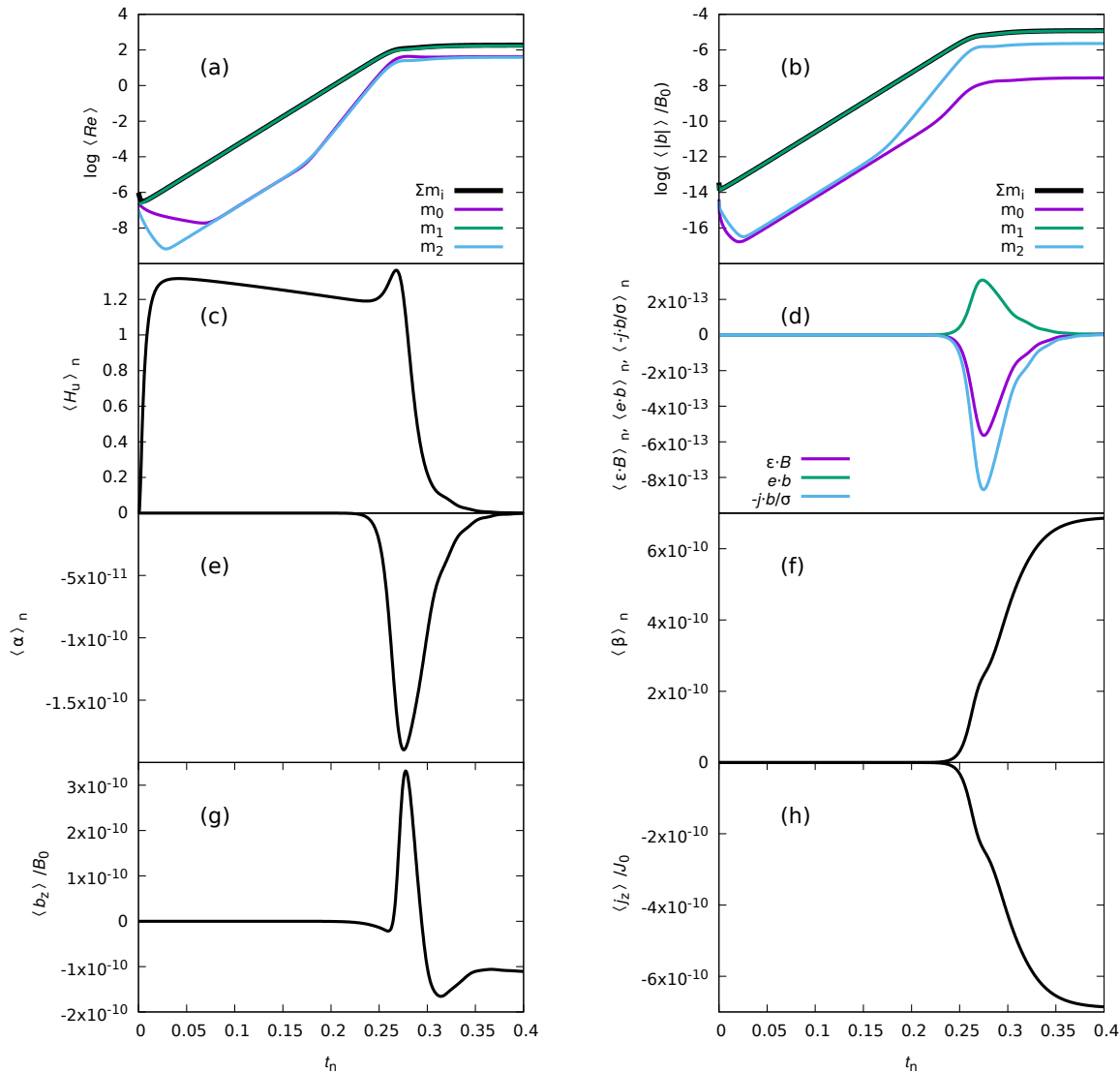}}
\caption{Time evolution of various quantities for $Pm=10^{-6}$ and
$Ha=60$. (a) - Reynolds number, (b) - Normalized Rms value of the induced 
magnetic field, (c) - normalized kinetic helicity, 
(d) - Relation of electromotive 
force and small-scale current helicity, (e) - Normalized mean 
$\alpha$ effect, (f) - Normalized mean $\beta$ effect, (g) - Normalized mean
axial field, (h) - Normalized mean axial current.}
\label{fig:fig2}
\end{figure}

Figure \ref{fig:fig2} exhibits the time dependence of various 
quantities that characterize the instability. 
The indicated dimensionless time $t_n$ 
is $t$ normalized by the viscous time scale 
$R^2/\nu$.
Figure 2a, to start with, shows the evolution 
of the averaged Reynolds number of the flow arising
from the initial state at rest:
$\langle Re \rangle=R/\nu (\langle u^2 \rangle)^{1/2}$ 
(here, $\langle...\rangle$ 
denotes an average over the total volume
rather than only over the azimuthal angle, which 
will always be indicated by an overbar).
We chose a logarithmic scale in order to 
evidence the exponential growth of the TI.
Approximately at $t_n=0.27$, the instability starts
to saturate.
We have added the respective energy
contents of the various azimuthal wavenumbers $m=0,1,2$. 
Evidently, the $m=1$ mode is the dominating one throughout
the evolution. However, approximately at $t_n=0.17$, both the
$m=0$ and $m=2$ modes start to increase with the double
growth rate as the $m=1$ mode. These even modes result from 
the non-linear term of the NSE. 
Saturation sets in 
when the $m=0$ and $m=2$ modes
have acquired an 
amplitude comparably to that of the
$m=1$ mode which ultimately brings the growth rate
of the TI to zero. The corresponding evolution of 
the averaged induced magnetic field is depicted in figure 2b.
Note that the $m=0$ component is here significantly 
weaker than the $m=2$ component, quite in contrast to the
rather parallel evolution for the kinetic energy.

The kinetic 
helicity 
$H_u= \bi{u} \cdot (\nabla \times \bi{u}) $
is the next quantity to be discussed (see figure 2c).
Actually, we show here
the helicity as normalized by the mean square of the
velocity, i.e. 
${\langle H_{u} \rangle}_n= \langle {\bi{u}} 
\cdot (\nabla \times {\bi{u}}) \rangle R /\langle u^2 \rangle $. 
After an initial increase,  this 
normalized helicity stays nearly 
constant for a while 
(i.e., the {\it non-normalized} helicity $\langle H_u \rangle$ 
grows with the same growth rate as the
kinetic energy), until it decays to 
zero when saturation is reached.

In Figure 2d we give an interpretation of the mean electromotive
force $\cal{E}$ in direction of the large scale magnetic field  
in terms of the small-scale current helicity $\overline{ \bi{j} \cdot \bi{b}}$, 
according to the relation
$\cal{E} \cdot \overline{\bi{B}}=-\overline{ \bi{j} \cdot \bi{b}}/\sigma+
\overline{\bi{e} \cdot \bi{b}}$ 
\cite{Bhattacharjee1986,Seehafer1994,Ji1999,Blackman2007,Ebrahimi2014}.
We see that this relation is perfectly fulfilled, showing that
$\alpha$ is essentially proportional to the current helicity, with 
a minor correction coming from an electric field term
(note that all quantities are normalized here by $B_0 J_0/\sigma$).

Figure 2e depicts separately the $\alpha$ effect, 
defined by $\alpha=\overline{({\bi u} \times \bi{b})} 
\cdot {\bi{B}}_0/B^2_0$. 
Since $\alpha$ has the dimension of a 
velocity, we give it here in form of a 
magnetic Reynolds number which includes again a complete 
spatial 
average: ${\langle \alpha\rangle}_n=\mu_0 \sigma R \langle 
({\bi u} \times {\bi{b}}) \cdot {\bi{B}}_0 \rangle /B^2_0$.
We observe initially an (exponential) increase, though to
a very small value in the order of $10^{-10}$, and then a 
decay to zero.

More monotonic than  $\alpha$, 
is the time evolution of 
$\beta=\overline{({\bi u} \times \bi{b})} \cdot {\bi{J}}_0/(\mu_0 J^2_0)$ effect
(figure 2f), 
which we normalize here by the
magnetic diffusivity $(\mu_0 \sigma)^{-1}$. This normalized, and 
spatially averaged 
${\langle \beta \rangle}_n=  
\mu_0 \sigma \langle ({\bi u} \times {\bi{b}}) \cdot {\bi{J}}_0 
\rangle/(\mu_0 J^2_0)$, 
acquires values of about $6 \times 10^{-10}$, so its
influence on the total resistivity can be considered 
as negligible.

The induction effects of the mean field coefficients $\alpha$ and 
$\beta$ are illustrated in figure 2g and 2h. Figure 2g shows the 
mean axial magnetic field $\langle b_z \rangle$ 
which is produced by the azimuthal
current that is driven, in turn, by $\alpha$. Normalized to $B_0$, 
we see again that this induction is negligibly small. 
Note also  that $\langle b_z \rangle$, in contrast 
to $\alpha$ (figure 2e), does not completely vanish in the 
saturation regime. We attribute this to numerical
inaccuracies which seem to ``stray'' some energy from the much stronger
$m=1$ mode into the $m=0$ and $m=2$ modes, an effect that is already 
visible in the non-physical parallelism of all  modes in the 
beginning of the exponential growth regime (figure 2a).
In contrast to this slight discrepancy, the behaviour of  
$\langle j_z \rangle/J_0$ (figure 2h) 
is nearly identical to that of $\beta$  (figure 2f).

\begin{figure}[h]
\centerline{
\includegraphics[width=0.9\columnwidth]{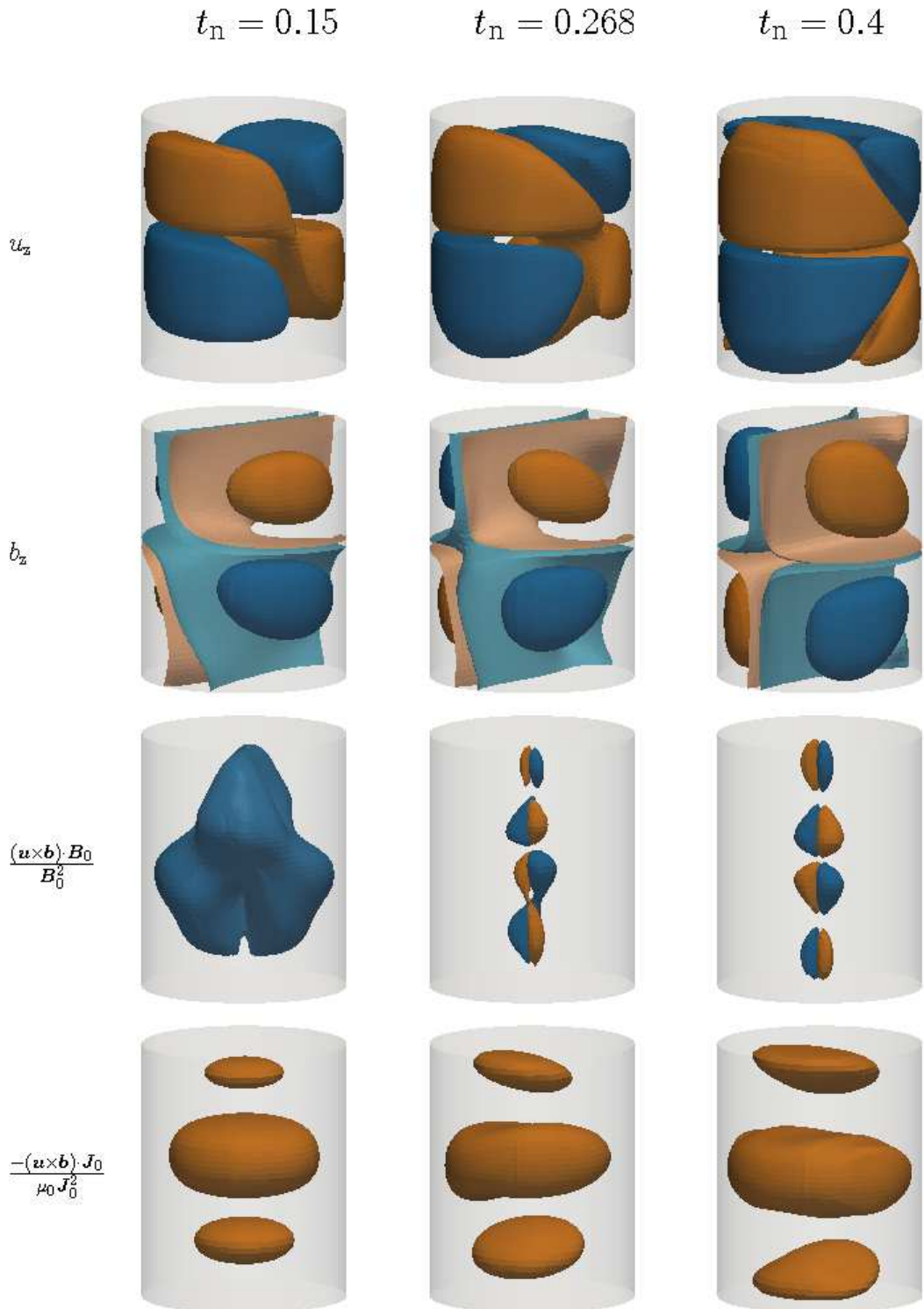}}
\caption{Snapshots of different quantities for three 
different instants. The contours correspond to 
7\%, 3\% and 50\%, 20\%, and 10\% of the
extremal values in the first, second, third and 
fourth row, respectively.}
\label{fig:fig3}
\end{figure}

For three selected instants in time, figure \ref{fig:fig3} 
illustrates the spatial structure of various features of the
TI (the normalized values of 
$u_z$, $b_z$, $(\bi{u} \times \bi {b})\cdot {\bi B}_0$, 
$(\bi{u} \times \bi {b})\cdot {\bi J}_0$). The left column depicts these 
quantities amidst the
exponential growth phase  in which we observe a clear
helical structure. The middle and right columns show 
then the respective structures shortly before and 
during the saturated regime.
Evidently, the helicity has completely disappeared here.
In terms of the mode structure, we notice that the
left handed spiral ($m=1$, say) and the right
handed spiral ($m=-1$, say) 
have grown to the same strength.

\begin{figure}[h]
\centerline{
\includegraphics[width=0.9\columnwidth]{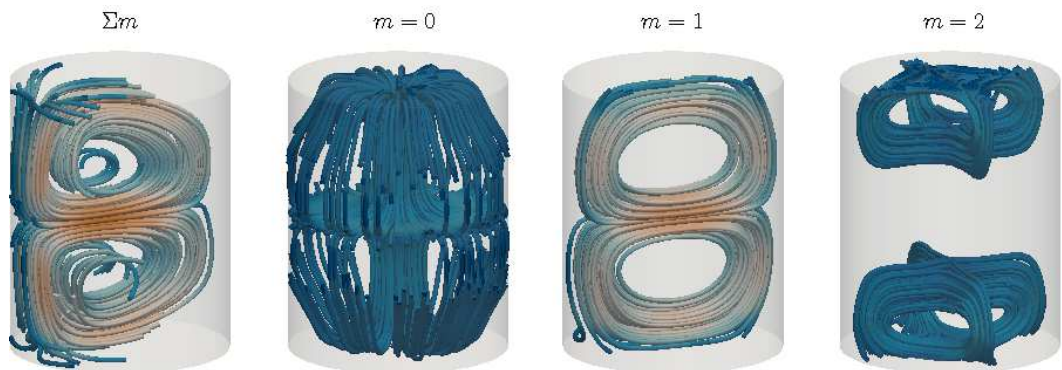}}
\caption{The velocity field in the saturated state, including 
the three lowest azimuthal modes. 
}
\label{fig:fig4}
\end{figure}

Having seen that, due to the low values of $Pm$, neither 
$\alpha$ nor $\beta$ is able to induce any relevant change
of the electromagnetic base state that would lead to
saturation of the TI, we have to look  for an alternative 
saturation mechanism. Evidently, this can only be related 
to a change of the hydrodynamic state, i.e. the flow field. 
Let us return to figure \ref{fig:fig2}a: after a long 
exponential growth period of 
the (more or less) pure $m=1$ mode, 
at $t_n=0.17$ the $m=0$ and $m=2$ modes start to grow due
to the action of the nonlinear terms in the NSE. 
As shown in figure \ref{fig:fig4}, the $m=0$ part of the 
velocity in the saturation comprises two poloidal
vortices pointing outward in the equatorial plane (in the 
dynamo community this flow topology would be denoted  
by s2$^+$ \cite{Dudley1989}). This axisymmetric
state, together with the $m=2$ component, 
changes the hydrodynamic base state for 
the TI so that it becomes just marginally stable.

With the simultaneous appearance of an $m=0$ and $m=2$ component, 
it is no surprise that the saturation is also connected
with a restoration of the chiral symmetry.
According to the sum rule for the nonlinear interaction 
the $m=2$ will produce, from any dominant
$m=1$ mode, a corresponding $m=-1$ mode, so that
chiral symmetry is ultimately restored.

\subsection{Saturation with helicity oscillations}

\begin{figure}[h]
\centerline{
\includegraphics[width=0.99\columnwidth]{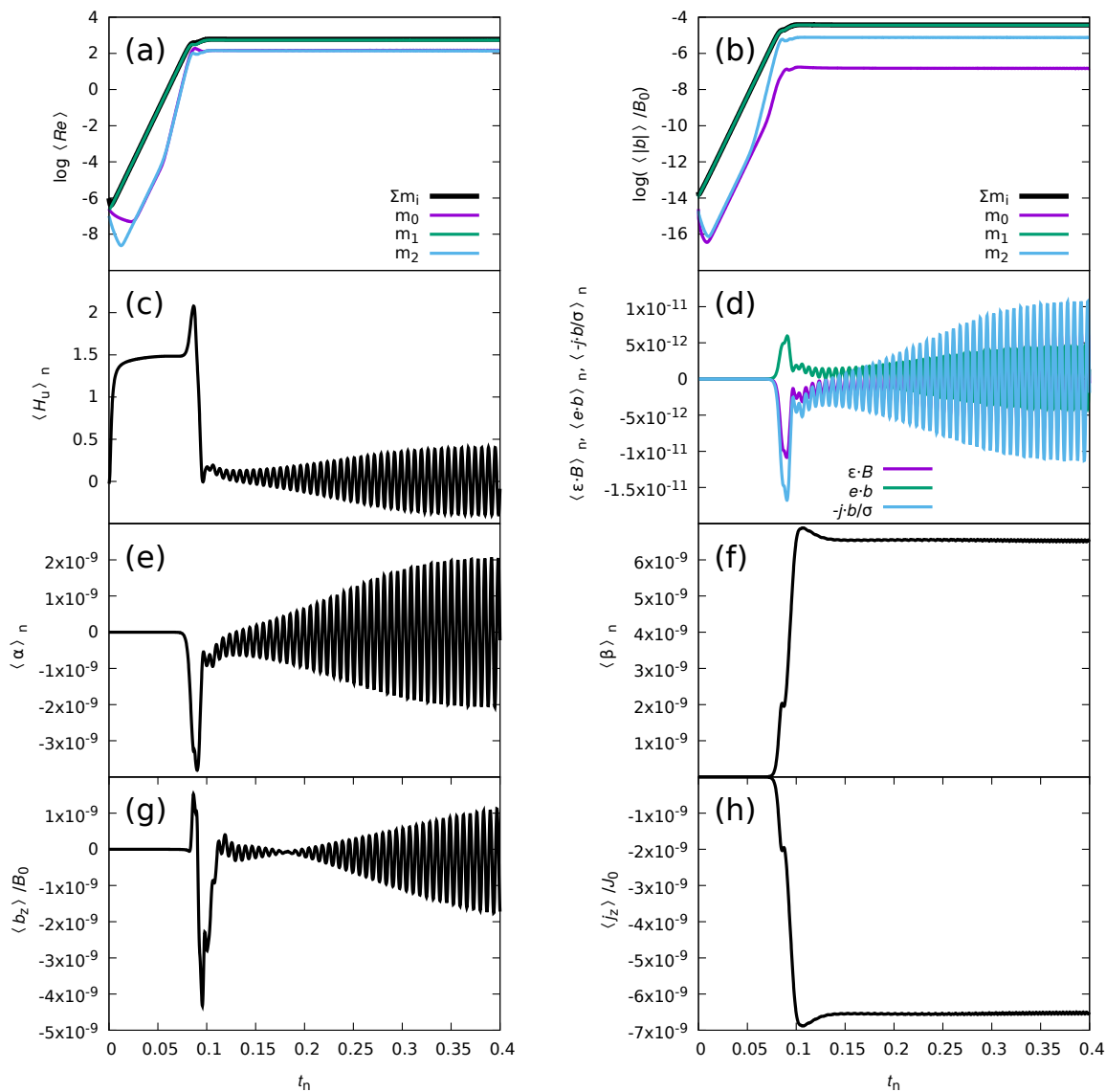}}
\caption{Same as figure \ref{fig:fig2}, but for $Pm=10^{-6}$ and 
$Ha=100$.}
\label{fig:fig5}
\end{figure}

We now increase the Hartmann number from $Ha=60$ to $Ha=100$.
As before, figure \ref{fig:fig5} illustrates the time evolution
of various quantities. While the behaviour of the 
Reynolds number and the $\beta$ effect
are not significantly different from the previous case
with $Ha=60$, the helicity and $\alpha$ look quite differently.
Evidently, the TI does not anymore saturate with 
zero helicity, but instead produces a helicity 
oscillation in the final state. Figure 6 shows again 
the spatial structure of various quantities in the 
exponential growth phase
and at two different instants in the saturated state. 
In the plots for 
$(\bi{u} \times \bi {b})\cdot {\bi B}_0$,
the helicity oscillation appears as a
slightly changing asymmetry between positive (orange) and 
negative (blue) ``blobs''.

\begin{figure}[h]
\centerline{
\includegraphics[width=0.9\columnwidth]{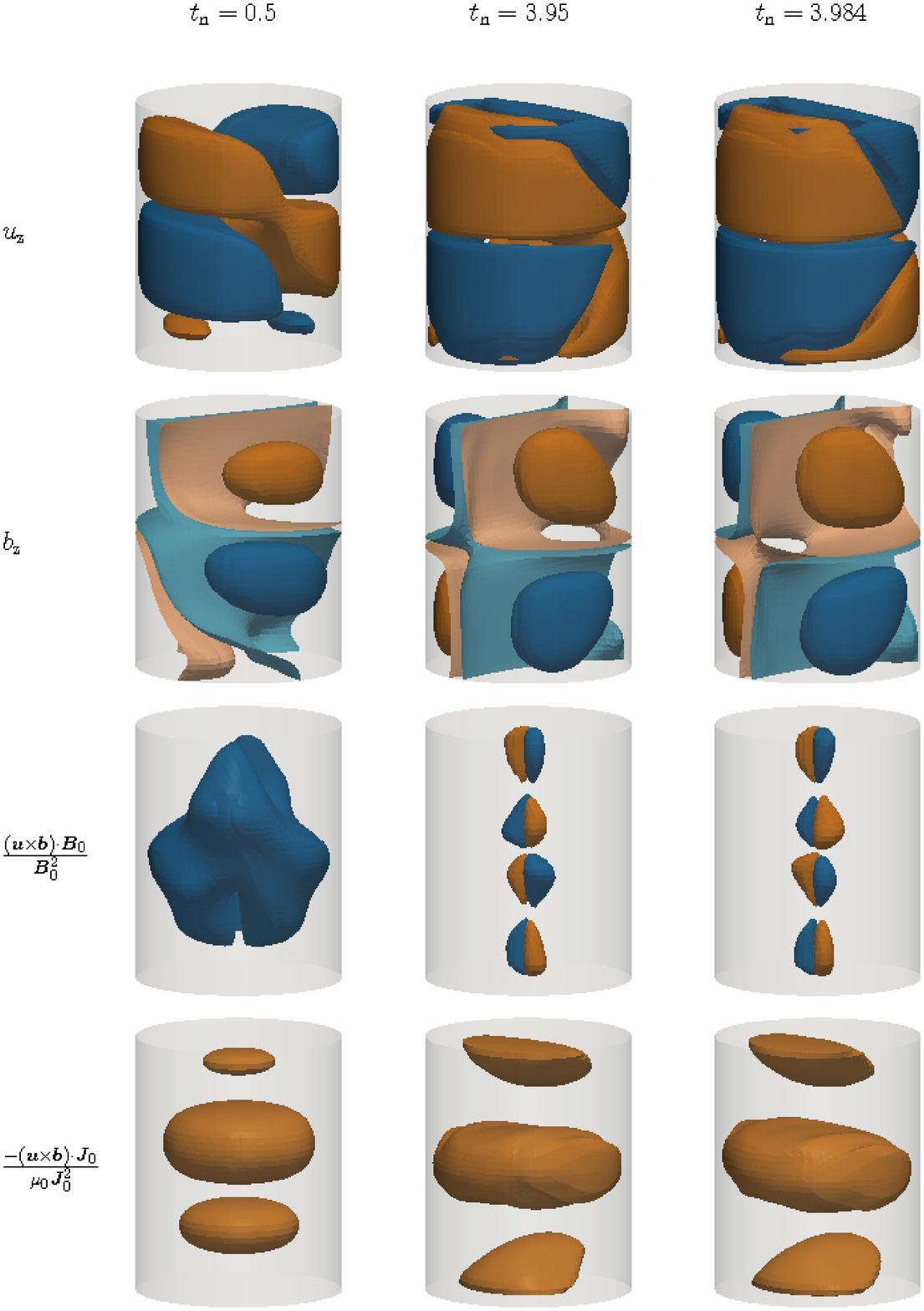}}
\caption{Same as figure \ref{fig:fig3}, but for $Pm=10^{-6}$ and 
$Ha=100$.}                        
\label{fig:fig6}
\end{figure}

\begin{figure}[h]
\centerline{
\includegraphics[width=0.90\columnwidth]{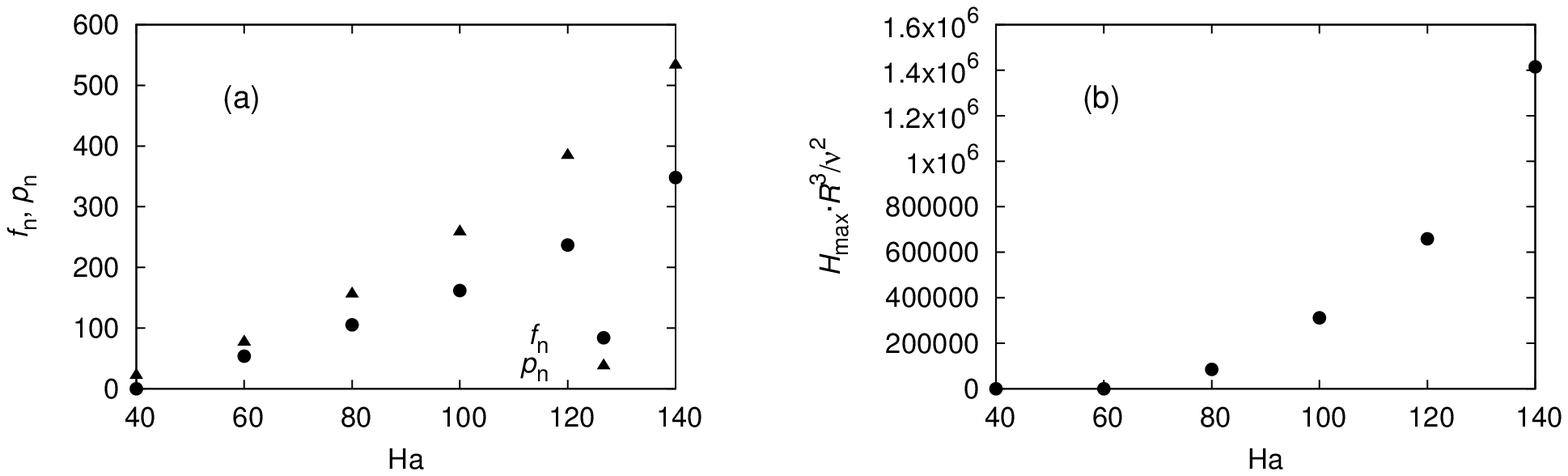}}
\caption{Characteristics of the helicity osciallations in 
dependence on $Ha$, for $Pm=10^{-6}$. (a) Frequency, also compared with the 
growth rate in the exponential growth phase. (b) Amplitude.}
\label{fig:fig7}
\end{figure}

For the range between $Ha=40$ and $Ha=140$ we characterize
this helicity oscillation by its amplitude and frequency
(figure \ref{fig:fig7}).

\subsection{Saturation with finite helicity}

We leave now the realm in which the magnetic induction is 
irrelevant for saturation and move towards the parameter region
which had already been explored by other codes. 
Actually, we increase the magnetic Prandtl number to
$Pm=10^{-3}$, and consider the case $Ha=100$.
This leads to a Lundquist number of $S=3.16$ for which we come close
to the edge of applicability of the inductionless approximation
\cite{Herreman2015}.

\begin{figure}[h]
\centerline{
\includegraphics[width=0.99\columnwidth]{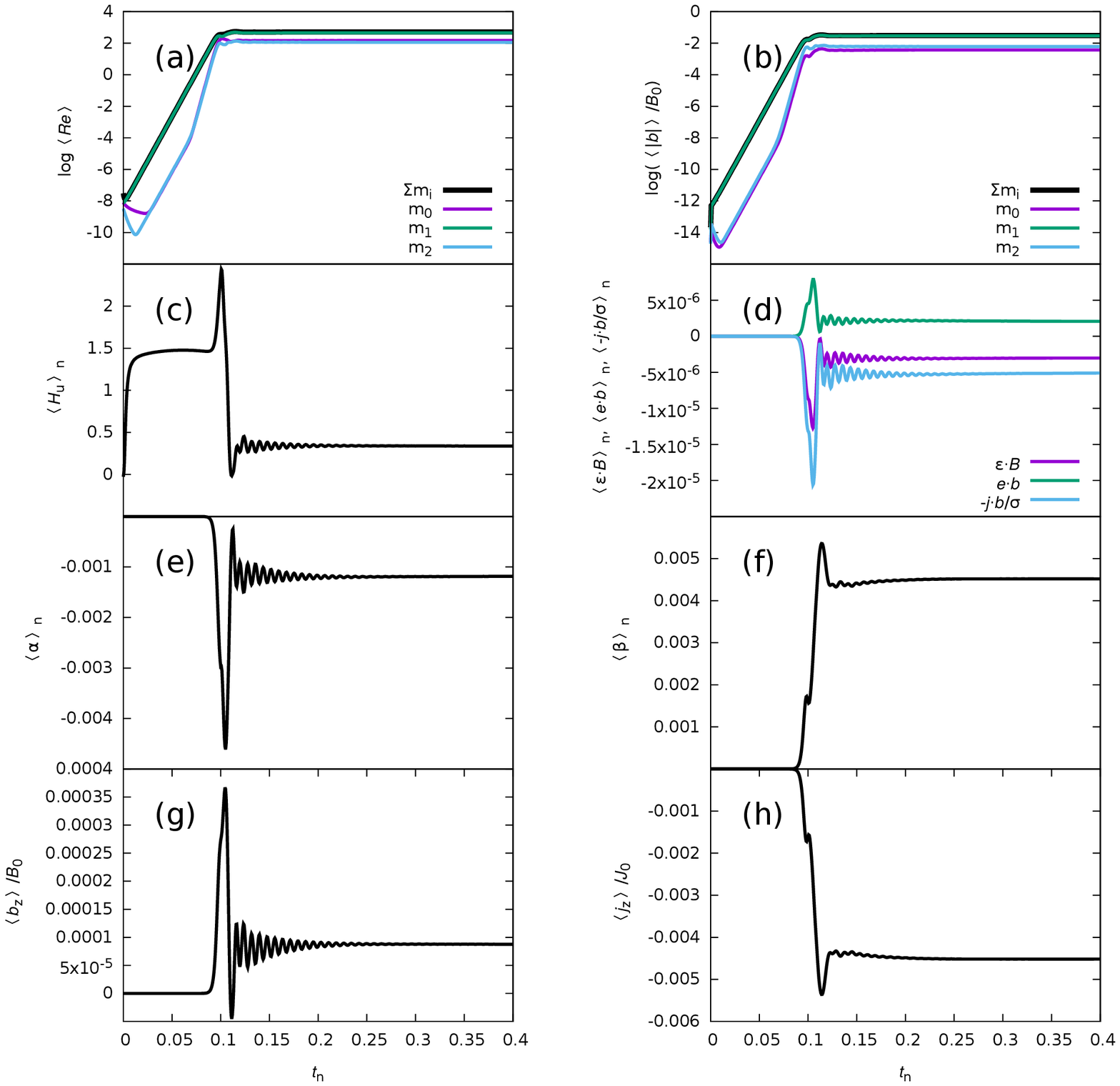}}
\caption{Same as figure \ref{fig:fig2}, but for $Pm=10^{-3}$ and $Ha=100$.}
\label{fig:fig8}
\end{figure}

\begin{figure}[h]
\centerline{
\includegraphics[width=0.9\columnwidth]{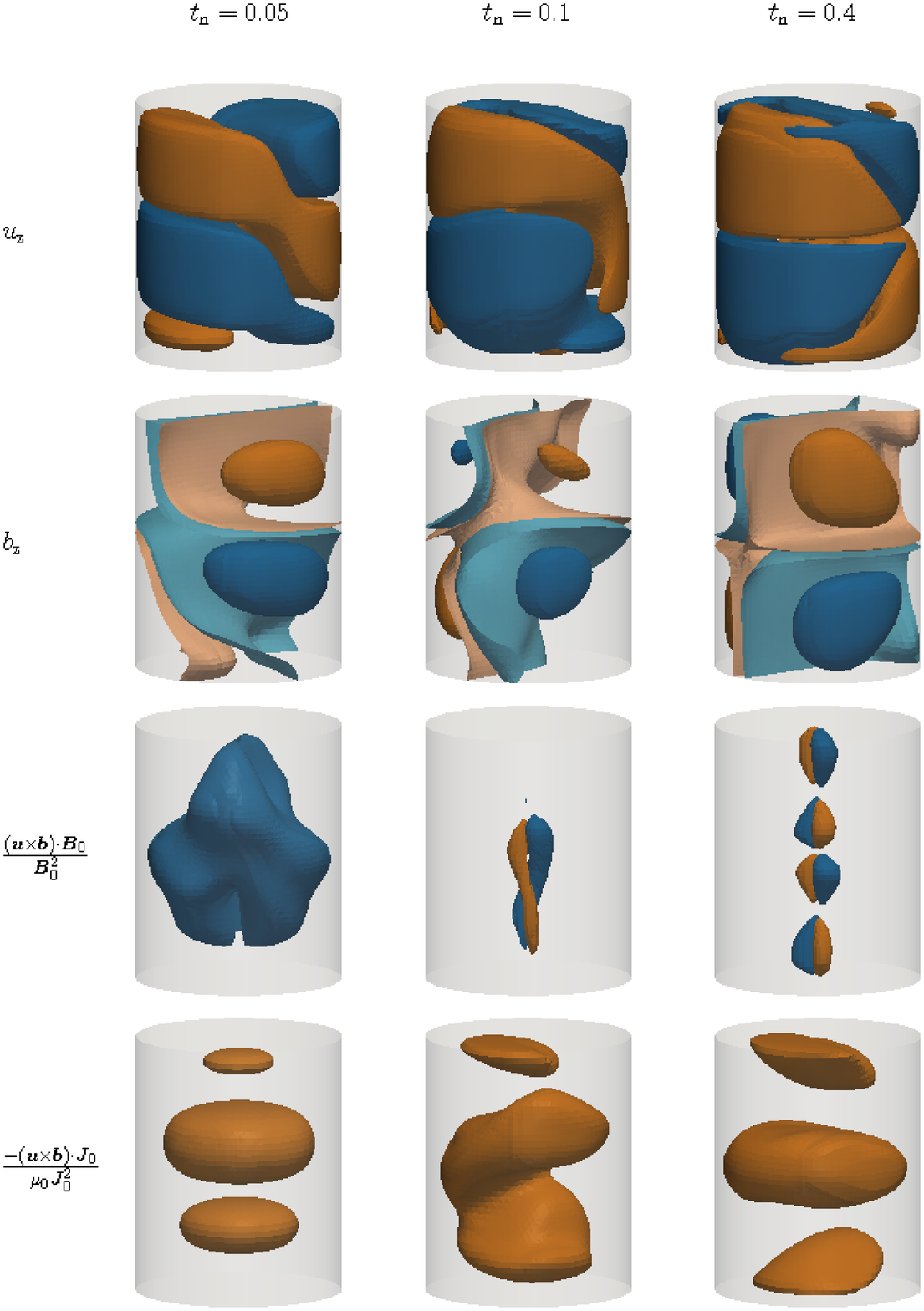}}
\caption{Same as figure \ref{fig:fig3}, but for $Pm=10^{-3}$ and $Ha=100$, and
different instants.}
\label{fig:fig9}
\end{figure}

Again, figures \ref{fig:fig8} and \ref{fig:fig9} show the 
time evolution, and some 
snapshots, of various quantities.
The main difference to the former runs at low $Pm$ is that
now the helicity and $\alpha$ acquire non-zero values in the
saturated state. This is also shown in the snapshots of figure \ref{fig:fig9}.
Evidently, we are now in a regime where 
the induced magnetic fields
contribute significantly to the saturation mechanism.
This involves also that the $m=0$ component of the magnetic field
(\ref{fig:fig8}b)
becomes now comparable to the $m=2$ part, quite in contrast
to the former cases with $Pm=10^{-6}$.

\begin{figure}[h]
\centerline{
\includegraphics[width=0.9\columnwidth]{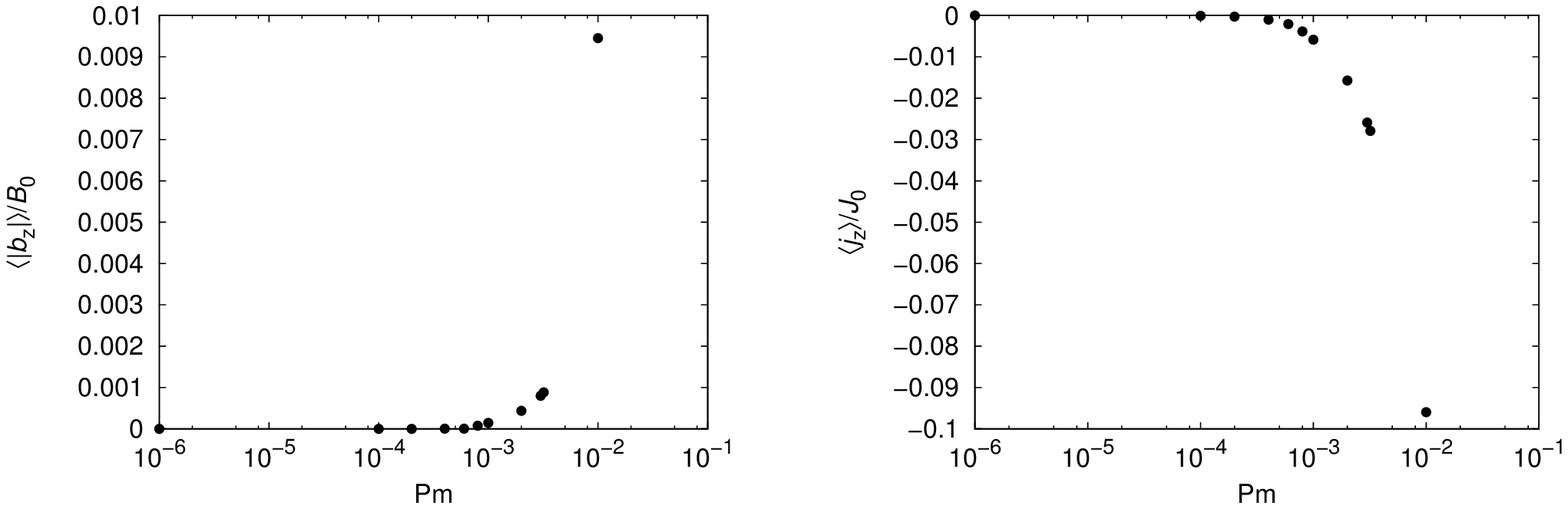}}
\caption{Dependence of 
$\langle b_z \rangle/B_0$ and $\langle j_z \rangle/J_0$
on the magnetic Prandtl number}
\label{fig:fig10}
\end{figure}

This is illustrated in figure \ref{fig:fig10} which shows the dependence 
of the induced mean current $\langle j_z \rangle/J_0$ 
and the induced mean axial field $\langle b_z \rangle/B_0$
on $Pm$ (at fixed $Ha=100$). According to the criterion of 
Kruskal-Shafranov \cite{Bergerson2006}, we know that
$b_z$ tends to inhibit the TI, so that a finite value of 
$\alpha$ is indeed likely to appear in the saturation regime.

\subsection{Between chiral symmetry breaking and helicity oscillations}

In the following, we will summarize the three different saturation 
mechanisms. We will be guided by the 
simple and instructive model of chiral symmetry breaking model 
that had been worked out 
by Bonanno \etal \cite{Bonanno2012}.

The  authors started from left and right handed helical modes
for the velocity and the magnetic field, 
fulfilling the Beltrami relation 
\begin{eqnarray}
\nabla \times {\bi L}=\lambda {\bi L} \;\;\; \mbox{and} \;\;\;
\nabla \times {\bi R}=-\lambda {\bi R} 
\end{eqnarray}
which can be realized by appropriate linear 
combinations of Chandrasekhar-Kendall functions  
$J_m(r \sqrt{\lambda^2 +n^2\pi^2/h^2}) \cos(m\phi) \cos(nz\pi/H)$.

Invoking some symmetry arguments, 
the authors ``guessed'' 
the simplest form of a Lagrangian which then
leads to the following evolution equations for
the energy of the left and right handed helical modes

\begin{eqnarray}
\frac{d E_L}{d t}&=&2 \gamma E_L -4 \mu E_L^2-4\mu_{*} E_L E_R\\
\frac{d E_R}{d t}&=&2 \gamma E_R -4 \mu E_R^2-4\mu_{*} E_L E_R \; .
\end{eqnarray}

The last terms on the r.h.s.~of equations (6,7) describe the 
so-called mutual antagonism between the two chiralities 
that has been used extensively
in the theory of
homochirality of bio-molecules 
\cite{Frank1953,Brandenburg2005,Saito2013}.

From here, one arrives at the following evolution equations for the
total energy $E=E_R+E_L$ and the helicity $H=E_R-E_L$:

\begin{eqnarray}
\frac{d E}{d t}&=&2 \gamma E -2(\mu+\mu_{*})E^2-2(\mu-\mu_{*})H^2\\
\frac{d H}{d t}&=&2 \gamma H -4\mu E H  \; .
\end{eqnarray}

\begin{figure}[hbt]
\centerline{
\includegraphics[width=0.7\columnwidth]{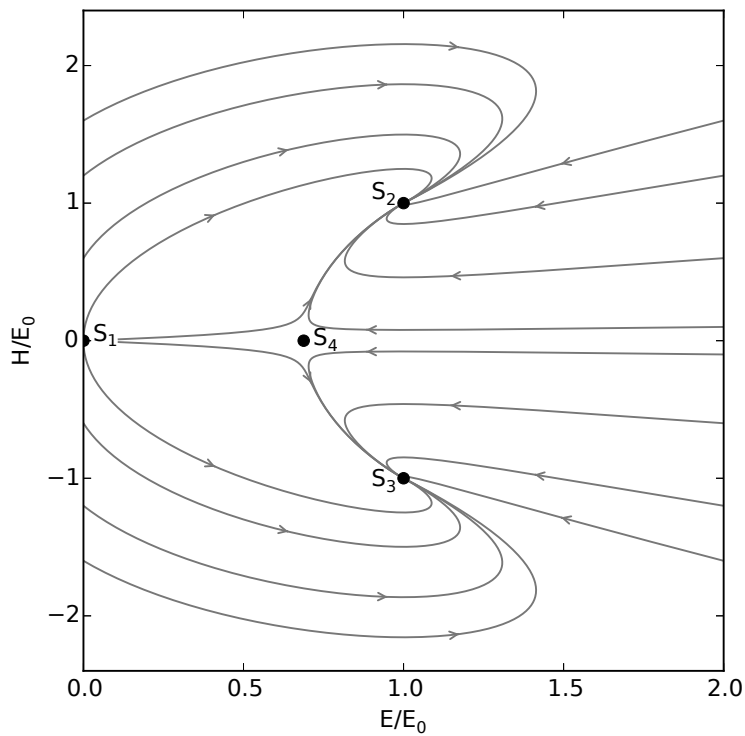}}
\caption{Phase portrait of the coupled equation system 
(8,9), for the parameters $\gamma=2.71$, $\mu=3.0$, 
$\mu_{*}=5.7$. In this typical situation, $S_1$ is a 
repeller, $S_2$ and $S_3$ are 
attractive points, while $S_4$ is a saddle point.}
\label{fig:fig11}
\end{figure}

In figure \ref{fig:fig11} we illustrate the phase portrait of this 
equation system showing a clear chiral symmetry breaking 
both in the exponential growth
phase as well as in the saturated phase.

\begin{figure}[hbt]
\centerline{
\includegraphics[width=0.9\columnwidth]{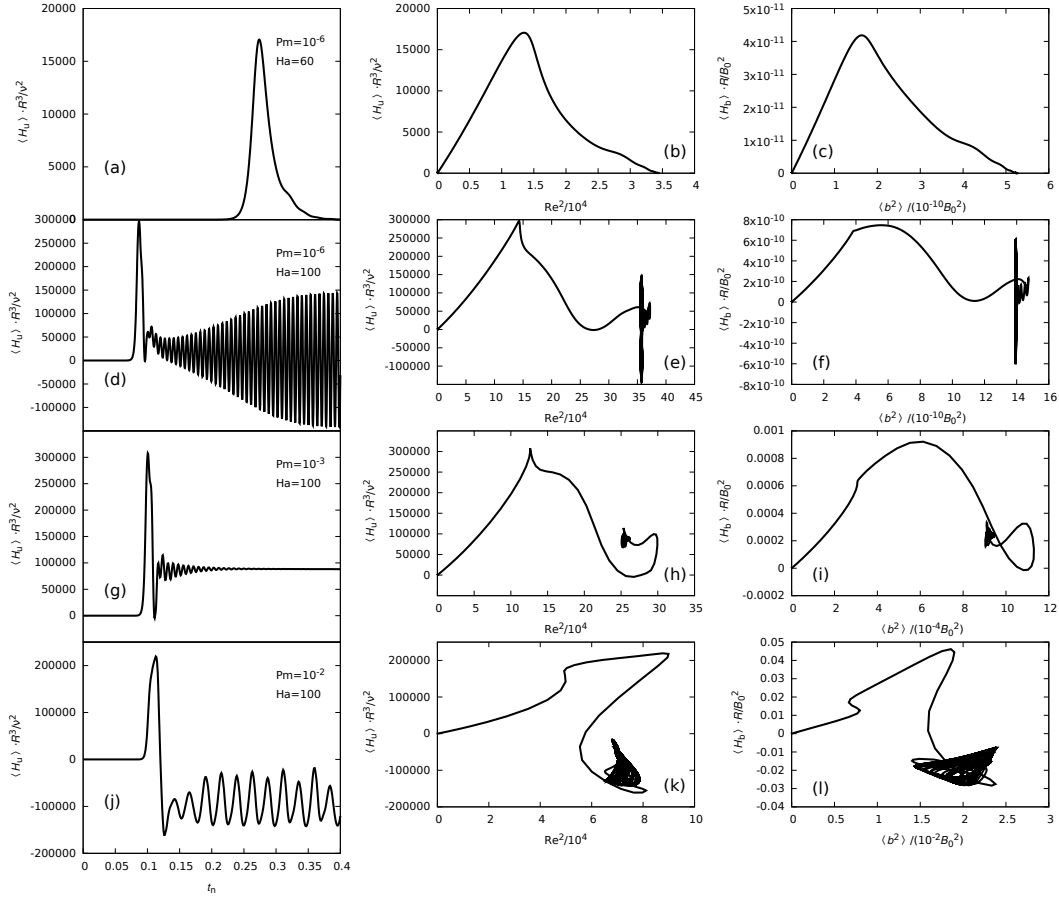}}
\caption{Time evolution (left), and phase portraits 
for the kinetic helicity (middle), and phase 
portrait of the current helicity (right),
for four different cases. Note that the viscous time scale
is indeed the relevant one both for the growth process and for
the helicity oscillations.}
\label{fig:fig12}
\end{figure}

We return now to our three cases, for which we 
plot the time evolution (Figure \ref{fig:fig12}, left columns) 
and the phase portrait  
of the kinetic helicity (middle column), as well as the
phase portrait of the current helicity (right column).
To make contact with figure 11, we have now chosen a
different normalization so that both helicities
start at zero.

We start, in the first row of figure \ref{fig:fig12}, with  
$Pm=10^{-6}$ and $Ha=60$.
Evidently the exponential growth phase looks very similar as in
figure 11, but ultimately the system runs into 
a state with zero helicity.

The second case with $Pm=10^{-6}$, $Ha=100$ is similar but terminates with
a helicity oscillation around zero. It is remarkable  that this
helicity oscillation proceeds without any significant oscillation
of the energy. 

The third case, $Pm=10^{-3}$, $Ha=100$ has indeed some resemblance 
with the above model 
of Bonanno \etal and terminates with a finite, non-zero helicity.

In the fourth row, we add here also the plot for $Pm=10^{-2}$ and $Ha=100$
which amounts to $S=10$ which is beyond the simple applicability of
the quasistatic approximation and should be, therefore, considered with
caution. 
Evidently, the magnetic helicity shows now a phase portrait that is 
similar to figure \ref{fig:fig11}, despite the fact 
that we now observe an 
"overshoot" to the helicity with the other sign, and also
a remaining oscillation in the saturated state. While we do
 not claim that this is a reliable result, the two last rows of 
 figure \ref{fig:fig11} at least suggest 
 that we are now approaching the usual saturation scheme as already 
 discussed by \cite{Bonanno2012}.

\section{Conclusions} % Motivation

In this paper, we have utilized an integro-differential
equation solver for addressing the problem of chiral symmetry 
breaking in the exponential growth phase and 
in the saturation phase of the Tayler instability.
The advantage of this code is its easy applicability
for small magnetic Prandtl numbers, while
its suitability for problems at $S>1$ is at least 
questionable (at least it has to be checked case by case whether 
a final steady state with finite and non-oscillatory, i.e. static, 
helicity could still be treated with our quasistatic scheme). 
Our simulation have allowed 
to identify three different saturation regimes.   

To start with the last regime, for a comparably large value
of $S=3.16$ we have confirmed 
a similar type of chiral symmetry breaking as it was previously 
evidenced by Gellert \etal \cite{Gellert2011} and Bonanno 
\etal \cite{Bonanno2012}. Depending on the random initial
conditions, the TI grows with one of the two
possible helicities which does not disappear  
in the saturated regime.
The helicity is intrinsically connected with a non-zero
$\alpha$ effect that generates a current parallel to
the applied azimuthal magnetic field. At the same time
the mean-field e.m.f. 
contains also a significant  $\beta$ effect 
that changes the axial current. Both effects
together work against the TI. In the 
ultimate case of high $S$ (which is, probably, 
not accessible by our code) one could expect 
a sort of Taylor-relaxation into a 
helicity maximizing  state \cite{Taylor1986}.
Whether for those large values of $S$ one reaches 
a regime of helicity oscillation around a finite
value (as suggested by the fourth row of figure 
\ref{fig:fig12}) is still to be validated by 
complementary codes.

The saturation mechanism, which relies 
on changing (by mean-field induction effects) the 
electromagnetic base state 
in such a way that it becomes just marginally 
stable against TI, does not apply for $S\ll 1$. 
This can already be seen from the 
general scaling $Re\propto Ha^2$, or $Rm\propto S^2$ 
which applies both to $S<1$ and $S>1$ \cite{Weber2013}. 
For
$S\ll 1$ the final $Rm$ becomes much too small
in order to induce any significant changes of the
original applied magnetic field.

In this parameter region, the saturation 
relies instead on the non-linear appearance of an
$m=0$ and an $m=2$ velocity component,
which now changes the {\it hydrodynamic} base state of the TI in 
a way that the growth rate of the TI vanishes.

Perhaps the most interesting result of our study is
the observation of a robust and systematic helicity 
oscillation whose amplitude and frequency dependence 
on $Ha$ has been worked out.
Interestingly this 
helicity oscillation is not connected with 
any significant energy oscillations.

Based on the latter observation, we  would like to conclude 
with some, admittedly, very speculative ideas.  
There is a long tradition in trying to
link the various frequencies of the solar dynamo action to 
corresponding periodicities of planetary motion. Tracing 
back to a paper by Jose \cite{Jose1965}, who
had related the 11.86 years Jupiter orbit with the 22.08 
year solar cycle,
some refinements of this connection in  
terms of a combined torque and gravity action of  
Earth, Venus and Jupiter have been discussed recently \cite{Wilson2013}. 
Other papers have tried to link periodicities of the 
Jupiter-Saturn orbits to various 
longer-time cycles of the solar dynamo \cite{Scafetta2010,Abreu2012}, 
with possible connections to the climate on Earth. 
However, in all those cases, it was noticed that the planetary forces 
are much to weak to compete with the typical acceleration  forces
in the convection  zone. The only viable way of
influencing  the solar dynamo was speculated to rely on the
action of gravity on the shape, or local rotation rate, of the tachocline. 
Yet, this would imply that the solar dynamo  works indeed as
some sort of Tayler-Spruit dynamo \cite{Spruit2002}, 
in which the transformation from poloidal
to toroidal field is traditionally realised by 
differential rotation, the reverse mechanism, however, by  
some 
$\alpha$ effect due to the  TI.

It is exactly here where helicity oscillations, and 
their possible synchronization with planetary forces and torques,
might come into play. In particular, since the oscillations
of $\alpha$ are
not connected to any significant changes of the energetic content, 
very minor changes of the state of the tachocline might just open
the ``$\alpha$-bottleneck'' for the Tayler-Spruit
dynamo (which is, in any case, still a sort of 
$\alpha-\Omega$ like dynamo). Even if the $\alpha$ oscillations are 
around some non-zero mean values prevailing in the 
two solar hemispheres, it still might give rise to 
dynamo oscillations.

Note that a parametric resonance and synchronization 
of $m=1$ dynamo eigenmodes with $m=2$ 
velocity perturbations 
has been observed both for galactic dynamos 
(swing excitation, \cite{Schmitt1992}),
and for a VKS-like dynamo \cite{Giesecke2012}. 
Whether a similar effect may actually be at work for 
synchronizing the solar dynamo  with periodic
planetary forces via their action on the tachoclinic state, 
will remain a topic for future investigations.

\section*{Acknowledgment}

This work was supported by Helmholtz-Gemeinschaft Deutscher
Forschungszentren (HGF) in frame of the ``Initiative f\"ur mobile und
station\"are Energiespeichersysteme'', and 
in frame of the 
Helmholtz Alliance LIMTECH, as well as
by Deutsche Forschungsgemeinschaft
in frame of the SPP 1488 (PlanetMag). 
We thank Pascal Beckstein for his support
to increase the performance of our numerical
code. We gratefully 
acknowledge fruitful discussions with
Rainer Arlt, Alfio Bonanno, Axel Brandenburg,
Marcus Gellert, Wietze Herreman, Rainer Hollerbach, 
Caroline Nore, J\=anis Priede, 
G\"unther R\"udiger and Martin Seilmayer
on several aspects of the Tayler instability.

%LITERATURE
\section*{References}
%\bibliographystyle{unsrt}
%\bibliographystyle{iopart-num}
%\bibliography{database}
\providecommand{\newblock}{}

\end{document}